# Employing Wikipedia's Natural Intelligence

# For Cross Language Information Retrieval

Mikhail (Mike) Basilyan

Department of Computer Science
Johns Hopkins University
Baltimore, MD

13 May 2009

## Abstract

In this paper we present a novel method for retrieving information in languages other than that of the query. We use this technique in combination with existing traditional Cross Language Information Retrieval (CLIR) techniques to improve their results. This method has a number of advantages over traditional techniques that rely on machine translation to translate the query and then search the target document space using a machine translation. This method is not limited to the availability of a machine translation algorithm for the desired language and uses already existing sources of readily available translated information on the internet as a "middle-man" approach. In this paper we use Wikipedia; however, any similar multilingual, cross referenced body of documents can be used. For evaluation and comparison purposes we also implemented a traditional machine translation approach separately as well as the Wikipedia approach separately. (Directions for running the software are in Appendix A.)

## 1   Introduction

The goal of this project and paper is to demonstrate a new method for querying documents of one language in another. We use Russian as the target language and English as the query language, though the results and algorithms can be easily extended to any language combinations and even multiple languages. We introduce a new technique for creating a query vector in the target language (Russian) using Wikipedia, and then we combine that vector with the vector derived using the machine translated query. From here on we will refer to the Wikipedia approach as such, and to the machine translation method as either the traditional method, the current method, or the "Babelfish method" after Yahoo!'s "Babelfish" translation tool.

### 1.1   Cross Language Information Retrieval

Cross Language Information Retrieval (CLIR) is the process of retrieving documents in a language or languages other than that of the original query. CLIR is important for many applications. For example, there are many document collections that are made up of multiple languages and it's more convenient to search the entire collection with the same query. Another example is the process of searching for an image, graph, or diagram that may be a part of larger document. If, for example, we are looking for a circuit schematic for some electronic device, it might not matter to us that the schematic is part of a document written in a language we cannot understand.

## 1.2  Current Methods

Current methods of CLIR usually involve a machine translation (MT) step. The query, the documents, or both are translated with a MT algorithm and compared using traditional information retrieval techniques. This process is limited by the effectiveness of modern MT algorithm. The technique proposed here and described below attempts to side step the issues of MT. Its advantages are described below.

## 1.3  Using Wikipedia for CLIR

Wikipedia[i], the free online encyclopedia, has become an invaluable resource for many individuals looking for fast information. It is probably one of the biggest collections of freely available, organized human knowledge to date and contains millions of articles on obscure topics in a wide variety of languages: currently almost 3 million in English, about 800,000 in French, and approaching a 900,000 in German. Articles on a variety of topics are written in multiple languages, though they are not translations of each other. For example, an article on "Golden Gate Bridge" appears in both English, and Italian.[ii] Even though the Italian article is not a direct translation of the English, it is very near to it in terms of fulfilling an informational need. Someone trying to find information on the famous red bridge in San Francisco would be similarly satisfied with either article, assuming they understand both languages.

The novelty of our approach lies in the use of Wikipedia's articles in different languages to transpose our informational need from the vector space in one language into another. For example, in order to search a French document collection for the English phrase "Big Ben," it is possible to perform a Wikipedia search for an English article closest in vector space or otherwise related to the term "Big Ben." Then using the contents of the cross referenced page in our target language, French, as a search query, we can identify the French documents in our target collection closest to it in vector space. Figure 1 illustrates this example graphically.

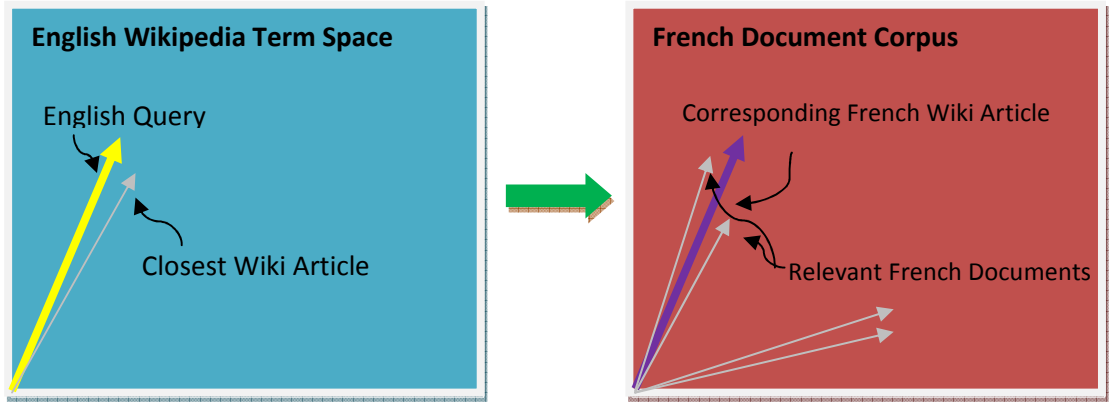

Figure 1: Using Wikipedia to move from one language space into another

## 1.4 Added Advantage

Using Wikipedia to map a language query into the desired space provides us with a number of advantages over traditional CLIR techniques.

- Machine translation algorithms are not available for a variety of the world's less commonly spoken languages. Wikipedia however provides articles in hundreds of languages, with 88 languages containing over 10,000 articles and 25 languages with 100,000 articles. In comparison, using the Yahoo! Babel Fish service[iii] can translate from English into just 12 other languages. From Spanish to just 2 others, and from Japanese to 1 other language.
- We are also able bypass the traditional problems associated with machine translation. For example translating "Big Ben" using the Yahoo!'s Babelfish service[iv] into Russian yields: "Большое Бен" which is a literal translation with a grammatical mistake and would yield very few hits. The common way to refer to "Big Ben" in Russian is, "Биг-Бен,"[v] this is more of a transliteration. (Wikipedia gets it right.)
- As Wikipedia grows and includes more obscure articles, this technique will become more effective.
- This technique is not limited to Wikipedia but can be used with any translated and cross referenced document collection that can be used as "middle-man."

## 1.5 Combining the Two Methods

We implemented a CLIR search engine that has three options for searching a document in a foreign language. The first method allows the user to use the machine translated query in combination with the Wikipedia derived query to search the document space and get the best of both approaches. We also implemented the Wikipedia approach and the machine translation approach (using Yahoo!'s Babelfish machine translator service) separately for comparison purposes.

## 2 Algorithms, Methods, & Tools

### 2.1 Languages & Libraries

Perl was chosen as the language of choice for implementing the algorithms. Perl's flexibility and extensibility make it a language of choice of rapid experimentation. Its text manipulation and regular expression facilities also help make it a good choice.

In addition to Perl and its built-in modules, the following modules were used: Google::Search for searching Wikipedia for the English articles, HTTP::Request, HTTP Response, LWP:: UserAgent for retrieving Wikipedia pages and parsing the URL for the target language (Russian) from them. Lingua::Stem::Snowball was used to stem the Russian words, Lingua::Translate[vi] was used as the machine translation algorithm with which to compare our results.

### 2.2 The Wikipedia Approach Algorithm

Figure 2 shows the outline of the overall process flow for the Wikipedia approach. The backend loads the document corpus (right hand column), removes common words (using a Russian stoplist) and stems the words using the Snowball stemmer[vii]. Then each document is encoded into its own vector using TF-IDF. On the user-end side, an English query is prompted, the closest matching Wikipedia page is retrieved, and the corresponding Russian page is fetched and is encoded into a vector. Finally, each document vector is compared against the Russian Wiki page vector and the results are returned, sorted.

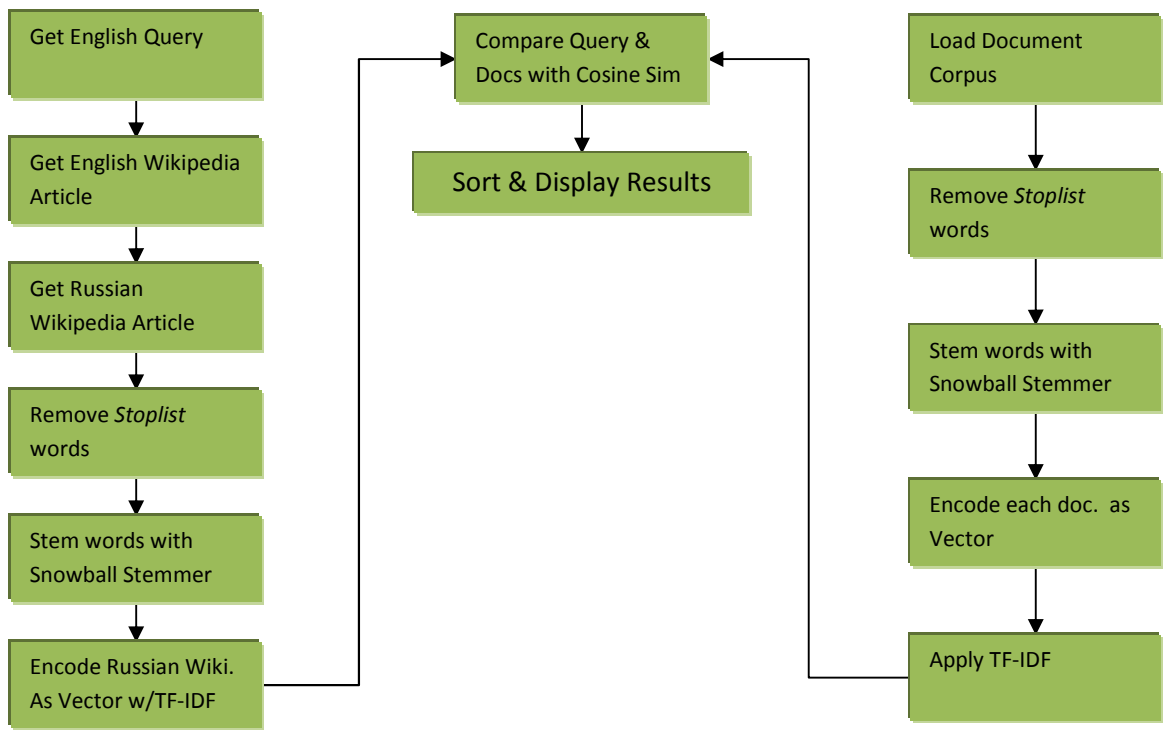

Figure 2: Wikipedia approach

### 2.2.1 Finding Corresponding Wikipedia Articles

Finding the English Wikipedia article is done using the Google search API. We use Google to search the Wikipedia site for the English using the query *"$query site:wikipedia.org;"*. It is important to point out that we are not using Google to perform any of the actual searching of the target Russian document corpus. We are only using Google to find the corresponding Wikipedia English article. Even though Wikipedia provides its own search engine it proved insufficient for our purposes, and so Google was used.

Once the English page is found the link to the Russian equivalent is parsed out using regular expressions and that page is fetched.

### 2.2.2 Encoding Vectors

Once the Russian website has been retrieved it's parsed into individual words. Common words, also known as stop words, such as "Я" (Russian for "I"), "Вы" (Russian for "you") were removed. Then using the Snowball stemming algorithm we transformed words from their given tenses and conjugations into their most common form. For example, if a document or query contained the words "важно", "важное", "важной" (all variations of the word important) they would all be transformed into "важн."[viii] This helps make sure that someone searching for the word "важно" gets documents that contain "важной."

Each word was encoded into our vector using TF-IDF. For each word, we counted how frequently it occurs in the document and multiplied that by log(N/$df_t$) where N is the number of documents in the collection and $df_t$ is the number of documents the given term appears in at least once. TF-IDF is widely believed to be the best known weighting scheme in information retrieval. It increases both with the number of occurrences of a term in a document and it increases for the rarity of a term in a document.

### 2.2.3 Comparing Vectors

Document vectors were compared using the cosine similarity:

$$\cos(A, B) = \frac{A \cdot B}{||A|| * ||B||}$$

The cosine similarity represents the angle between two vectors, A and B in the term space. The closer the vectors are in the term space the more terms they share in common and are thus more closely related. Other advantages of using the cosine similarity include the fact that it's *self normalizing* so vector lengths are not taken into account.

## 2.3 Using a Machine Translation

Our machine translation was implemented by contacting Yahoo!'s Babelfish translator service and requesting that the English query be translated into Russian. The Russian query was then transformed into a vector the same way the Wikipedia page was.

## 2.4 Joining the two Approaches

The two queries were joined together by taking the union of the two vectors.

## 2.5  Document Collection

Because we could not locate an existing Russian document corpus for information retrieval purposes, we had to build our own by hand.  To this end we scrapped together on the order of one hundred paragraph- length segments from http://www.math-net.ru, a website that archives articles from Russian mathematical journals.  We primarily used articles from 'Дискретная математика' (Discrete Mathematics).  It is worth noting that the author's understanding of Russian and advanced discrete mathematics is embarrassingly low.

# 3  Evaluation & Results

We wanted to see what advantages, if any, the Wikipedia approach brings to the machine translation approach.  To test this, we used the two approaches separately (options 2, and 3 in the software menu) to run the codes on some discriminating queries that help illuminate the difference between these two approaches.

First let's take a look at some places for which the Wikipedia approach when used alone had failed.  We performed searches for the English query "complexity." This query helps demonstrate one of the main weaknesses of the Wikipedia approach: it turns out that there is no Russian Wikipedia article for "complexity" and so our search failed.  Trying the Babelfish/machine translation approach alone yielded a satisfactory Russian translation: "сложность" and 15 apparently relevant results.

In a different trial both approaches performed equally well.  For example, searching for "monotonic functions" the results were surprisingly similar.

**With the Wikipedia approach:**

```
ENGLISH WIKI: http://en.wikipedia.org/wiki/Monotonic

TRANSLATED WIKI:
http://ru.wikipedia.org/wiki/%D0%9C%D0%BE%D0%BD%D0%BE%D1%82%D0%BE%D0%BD%D0%BD%D0%B0%D1%8F_%D1%84%D1%83%D0%BD%D0%BA%D1%86%D0%B8%D1%8F

RESULTS:
===============================================================================
RANK    DOC.ID    DOCUMENT TITLE              SIMILARITY
===============================================================================
01      0028      О средней сложности монотонных              0.153419
02      0088      Оценки сложности одного метода              0.153412
03      0081      О свойствах функций              0.134879
04      0012      О сложности расшифровки разбие              0.127879
05      0038      О мощности некоторых подклассо              0.117888
06      0061      Некоторые свойства групп инерц              0.110278
07      0044      Сложность умножения в некоторы              0.090094
08      0057      О числе биюнктивных функций              0.075312
09      0006      О связи уровня аффинности с кр              0.074252
10      0010      Независимые системы порождающи              0.071059
11      0086      Абстрактные свойства класса ин              0.062243
12      0032      Об уровне аффинности булевых ф              0.061459
13      0078      О числе независимых множеств в              0.058187
14      0039      Эквациональное замыкание              0.058107
15      0052      О наследовании свойств при суж              0.056978
```

**With the Babelfish approach:**

```
Please enter your English search query:
monotonic function
Contacting Yahoo!'s Babelfish Translator Service...
Babelfish Query Translation: монотонно функция
RESULTS:

==============================================================================
RANK    DOC.ID  DOCUMENT TITLE              SIMILARITY
==============================================================================
01      0028    О средней сложности монотонных          0.464489
02      0081    О свойствах функций         0.332347
03      0088    Оценки сложности одного метода          0.321211
04      0038    О мощности некоторых подклассо          0.317822
05      0061    Некоторые свойства групп инерц          0.261573
06      0012    О сложности расшифровки разбие          0.236581
07      0006    О связи уровня аффинности с кр          0.182960
08      0057    О числе биюнктивных функций     0.166212
09      0032    Об уровне аффинности булевых ф          0.151438
10      0036    О сложности вычисления диффере          0.129123
11      0052    О наследовании свойств при суж          0.122090
12      0025    Функция Шеннона сложности инте          0.114794
13      0024    О некоторых алгоритмах построе          0.110947
14      0003    язык Эта глава описывает одну           0.073595
15      0033    Бесповторность распознается сх          0.069425
```

The following test was meant to demonstrate areas in which the Wikipedia approach performed extremely well and the machine translation approach used alone failed. Take the query "bubble sort." Our document collection contains a paragraph long definition of the famous "bubble sort" algorithm that would be an ideal result for most queries for "bubble sort."

Using our Wikipedia method we get the following result:

```
==============================================================================
RANK    DOC.ID  DOCUMENT TITLE              SIMILARITY
==============================================================================
01      0000    Пузырьковая сортировка      0.194324
02      0023    Рандомизированный алгоритм мно          0.080642
03      0096    О сложности реализации конечны          0.069478
04      0003    язык Эта глава описывает одну           0.038339
05      0002    Распределение регистров     0.027259
```

The document we expected is ranked first (doc ID. 0000.)

Using Yahoo!'s Babelfish to perform the translation for "bubble sort" we get the translation "вид пузыря" which translates back into English as "a type of bubble." The expected document is not even amongst the top 7 results (our algorithm is configured to not return documents below a similarity threshold of 1e-12):

```
==============================================================================
RANK    DOC.ID  DOCUMENT TITLE              SIMILARITY
==============================================================================
01      0081    О свойствах функций         0.110099
```

```
02      0078    О числе независимых множеств в              0.081004
03      0061    Некоторые свойства групп инерц              0.072211
04      0059    Семейство многомерных статисти              0.064651
05      0073    О суммировании по путям в спек              0.064558
06      0068    О кодах Гоппы на одном семейст              0.051118
07      0024    О некоторых алгоритмах построе              0.045943
```

Running the above tests, amongst others, using software option 1 (combining the Wikipedia method with the machine translation method) yielded results that were as satisfactory as above or better. There are situations in which the Wikipedia approach provides lower quality results, especially when the article in question has many marginally related topics. In these situations it is helpful to only use the first twenty words in the article which include, typically, the definition of the term. (This option is already in the software.)

# 4 Future Directions

This project has given very satisfactory initial results. It has demonstrated that it is possible to use Wikipedia to aid in cross language information retrieval tasks. In future revisions it would be valuable to consider other cross-referenced documents in addition to Wikipedia. Translated corporate websites come to mind.

It would also be valuable to find a more complete document corpus that could be used to evaluate the process as well as an expert group to determine which documents are relevant so that a more quantitative analysis can be performed.

Finally, it would be interesting to provide a mechanism that can learn which queries are helped by including the Wikipedia results and which are harmed by it and weigh the corresponding vectors appropriately, or allow the user to set the weights manually.

# 5 Conclusion

It appears that our approach provides significant improvements over the traditional machine translation method used alone. It is of no use in some situations when there is no Wikipedia article for the matching query or the query is too complex to be captured in a Wikipedia article. However, given an appropriate cross-referenced, multi-lingual alternative to Wikipedia, any query is reasonable, in theory. It performs extremely well in situations where you have small, difficult to translate queries such as "bubble sort" or "root locus." In conclusion, it seems to make sense that this approach is used in conjunction with machine translation algorithms, as implemented here, to augment difficult to translate phrases with Wikipedia derived "translations."

# Appendix A: Running the Code

The code requires the following modules to be installed from CPAN:

(The first three should come with a default Perl installation.)

HTTP::Request
HTTP Response
LWP:: UserAgent
Google::Search
Lingua::Stem::Snowball
Lingua::Translate

Then to run the code just type "perl main.pl" in the directory into which the search engine was unpacked. To run queries an active internet connection is required.